# The mechanics of glass and functionalised glass surfaces

E. Barthel, M. Beauvais, R. Briard, N. Chemin, D. Dalmas, C. Heitz, M. Klotz, P. Nael, A. Perriot, A. Pinquier, S. Roux, L. Serreau, E. Sondergard, D. Vandembroucq.

Surface du Verre et Interfaces,
Unité Mixte CNRS/Saint-Gobain
BP 135
93303 Aubervilliers Cedex
France

etienne.barthel@saint-gobain.com

## *Abstract*

Glass is highly sensitive to surface flaws generated by contact. Surface cracks threaten both its mechanical strength and visual aspect. Recent trends in glass functionalisation by grafting or coating lend an even more prominent role to the surface.
For these reasons, the glass surface mechanics, which couples surface physico-chemistry with mechanical response, must be considered in details to optimize glass processing and products. We review a few recent advances in the field of glass surface mechanics, with special emphasis on the local ductility of silicate glasses and the mechanical stability of films and coatings.

## *Résumé*

Les défauts de surface, si caractéristiques de la surface du verre, menacent à la fois sa résistance mécanique et parfois son aspect. De plus, la fonctionnalisation par adsorption ou greffage, qui s'est fortement développée au cours de ces dernières années, donne un rôle croissant à la surface.
C'est pourquoi la mécanique de surface, qui permet de comprendre le couplage de la physico-chimie de surface et de la réponse mécanique est un domaine clé pour l'optimisation des procédés et des produits verriers. Nous donnons quelques exemples d'avancées récentes que nous avons faites dans le domaine de la mécanique de surface du verre: outre les questions de réponse mécanique et de stabilité des films et revêtements fonctionnalisants, nous traiterons aussi de la part de la plasticité dans les déformations irréversibles de la silice amorphe et des verres.

## *Introduction*

Glass is so unique in its combination of transparency, mechanical stiffness and – nowadays - low manufacturing costs that despite its millenary history, it keeps very strong and sometimes hardly unchallenged positions on such markets as windows, windshields, displays, containers, etc. None of the other dominant technological materials of this century can claim to such an antiquity, except maybe wood.
Ironically, for glass, one single weakness threatens both key properties. Indeed, the propensity of glass to form surface flaws challenges both its mechanical resistance, resulting in a brittle material, and its visual aspect. Every suburb train rider, especially on sunny afternoons, has to admit from the ubiquitous window pane engravings that glass is easily scratched (Fig. 1). One even wonders what psychological impulse drives these suburb train artists to especially target the windows. One may surmise  that the staccato-like feel and painful screech of glass-scratching is a far more exciting experience than the dull soundless scratching of metals and plastics.

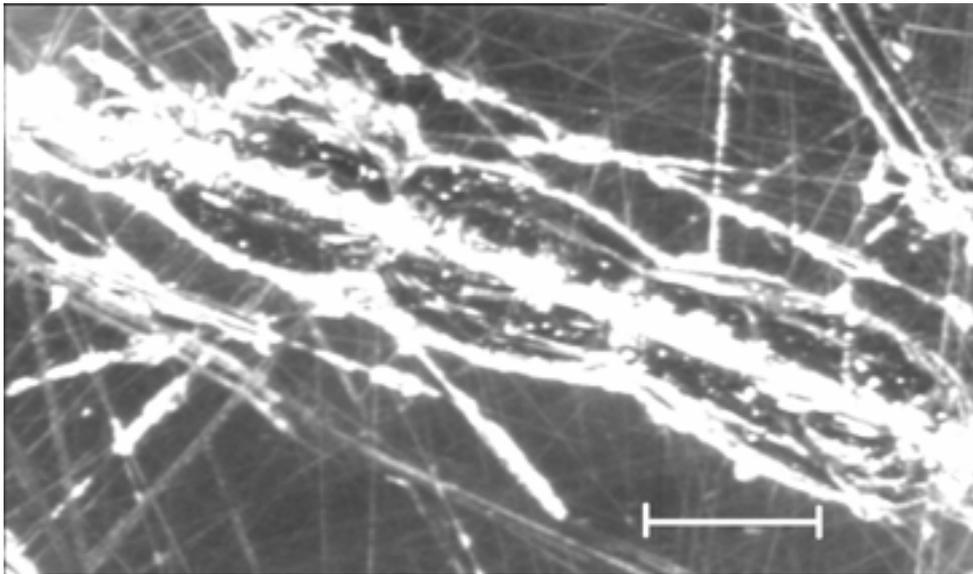

Fig. 1: Scratches on a glass floor tile. The bar is 50 µm. Photo G. Duisit, Saint-Gobain Recherche.

## *Glass surface and Surface Mechanics*

These differences in scratch behaviour reflect the contrast in mechanical properties. The technological fact that glass is brittle has long been known. Even the scientific concept that silicate glasses are the archetype of brittle materials can be traced back to the XVII[th] century. But scientific bases to understand glass brittleness came up only in the wake of the glorious scientific years of the late XIX – early XX[th] century, more precisely in the second half of the XX[th] century, when the specifics of the surface of materials are gradually recognized and the surface reaches a scientific status in its own right.

As an example, in 1930, Obreimoff measures the cleavage energy of mica in a controlled atmosphere [Obreimoff 1930]. His experiments pave the way to the fruitful combination of surface physico-chemistry and mechanics, or *surface mechanics*. And in 1934, when Derjaguin proposes a simple relation between interaction *energy* and interaction *forces* for curved bodies [Derjaguin 1934], he lays the foundations of the field of *surface forces* which is central for the present development of submicron structures and devices.

It is actually in the precedent decade that the earliest success of surface mechanics took place, when Griffith simultaneously elaborated the explanatory concepts for brittle fracture, based on the role of surface flaws, and backed them by experiments on glass [Griffith 1921].

Even this sound basis for the mechanisms of brittle fracture did not lead to the full mastery of glass surface flaws. Although strengthening strategies were devised and implemented, and the flaw density and their gravity reduced, it is fair to say that nowadays all glass products are still much more fragile than glass – perfectly flawless glass – in principle should be. But is there room for improvement ? One of the striking features is – as we show here - that the actual strain mechanisms in silicate glasses are not fully understood. In particular, glass may exhibit plastic behaviour. Now, in the formation process of surface defects, plastic strain is in competition with crack nucleation and growth. New insights in the small scale deformation mechanisms of glass, particularly plastic deformation, could be useful to reduce the sensitivity of the surface to flaw generation.

## *Glass functionalisation*

It must also be reckoned that presently, glass products rarely exhibit a bare surface. Among the glass surface modification techniques, some primarily aim at controlling the surface mechanics.

But a whole range of films or adsorbates has another purpose: imparting new properties to the materials. Indeed, the strong positions of glass among technological materials is also due to the fact that glass manufacturers have for many years been intent on bringing products with an unprecedented wealth of functions to the market. The recent bloom of the self-cleaning glasses is but one example.

Because the high processing temperatures involved in the glass manufacturing processes strikingly reduces the number of opportunities to impart original properties to the materials – despite the knowledge and skill of the glass scientists - one of the aptest ways to devise new functionalities for glass is to target the surface: surface modification can be carried out as a post-processing low temperature step, a sort of final touch. This opens a number of new avenues for improved optical, conduction, wetting properties for high technology glazings or displays for example. Other desirable properties target chemicals or particles exchanges between glass and its environment for filtration or contamination control.

Now once again, for all these products, mechanical requirements are crucial for their viability (Fig. 2): stiffness, durability, resistance.

What is the impact of these surface modifications and thin coatings on the surface mechanics of glass? We will show that this can be assessed through specific experimental tools and adequate modelling, so that one may eventually limit the additional weakness or enhance the mechanical improvements resulting from the functionalising coatings.

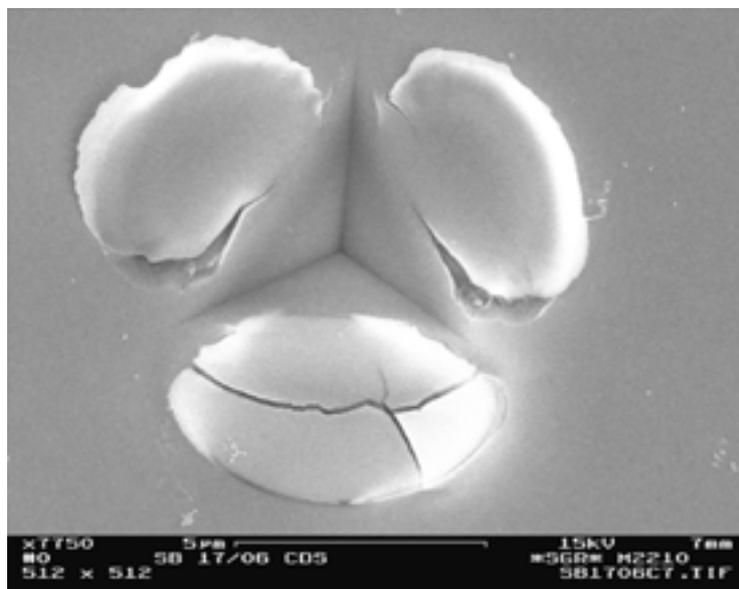

Fig. 2: indented sol-gel silica film on glass. The low adhesion and high yield stress of the film result in delamination of the film in the form of three flakes which generate highly visible optical defects on the film. Reprinted from E. Craven et al., Studies in Surface Science and Catalysis 146, 2003, 534 with permission.

## *Can plastic deformation help reduce surface flaws?*

People have dreamt for a perfectly plastic glass material for millennia: plastic glass is one of the many colourful inventions strewn around Petronius' Satyricon. In chapter LI, a glassmaker proudly presents the stunned emperor with a marvelous glassware, which, when thrown to the ground, does not shatter: the resulting bumps can even be mended with a hammer[1].

Indeed, brittleness actually results from a competition between plastic deformation and fracture. Hence for a given resistance to fracture (or toughness), a reduced brittleness may result from easier plastic deformation. Tempered steel is brittle...at the kilometer range [Rhee 2001].

Now glass does exhibit plastic deformations. The relevant lengthscale however is micrometric. Despite this fact, a singular feature of the ductile deformation of silicate glasses and most notably amorphous silica has been evidenced more than 30 years ago: it densifies [Ernsberger 1968]. In contrast to plastic deformation of metals, which occurs at constant volume, the plastic deformation of amorphous silica leads to *permanent increase* in density. Instead of deforming plastically by shear flow, silica deforms both by shear and by densification: the flow does not preserve the volume. The simplest evidence is that indented silica refracts visible light more than ordinary fused silica.

How can one assess these deformation processes? How can we propose a constitutive law for this phenomenon? The high pressures needed here can be applied by shocks. The pressure in diamond anvil cell experiments may also reach several tens of GPa, high enough to densify silica. But the stress field in shock experiments is ill-controlled. Diamond anvil cell experiments ideally result in a hydrostatic stress field, without shear components. Now densification and shear are coupled: this is quite unlike metals, where only shear determine the yield criterion, but is strongly reminiscent of granular materials like concrete. In such materials, shear stresses can *reduce* the density due to grain-grain decohesion. This is taken into account in the constitutive law (Fig. 3): in the pressure (p)-shear ($\tau$) plane, for granular materials, the plastic flow, at yield strength, goes backwards along the p-direction, leading to material expansion. In amorphous silica, it is found that shear *promotes* densification, presumably because it weakens the material structure. The slope of the yield criterion is then reversed: the plastic flow goes forward which results in densification [Lambropoulos 1996].

Thus for efficient modeling of indentation and scratch on such anomalous materials, experimental conditions combining hydrostatic stress and shear are required. This is the reason why we use indentation, which generates GPa stresses with a strong deviatoric component at the local scale, to assess the constitutive law.

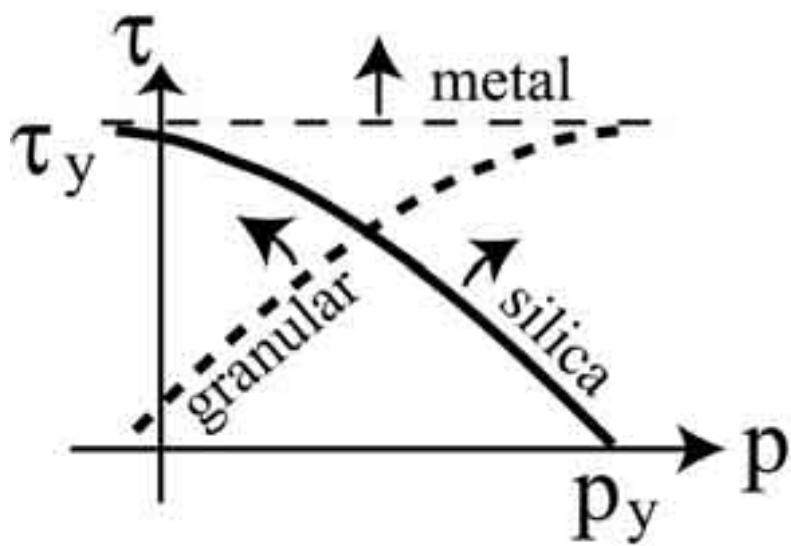

---

[1] For details on the enlightened marketing policy of this technological breakthrough, the reader is refered to the original text (T. Petronius, Satyricon, ch. LI).

Fig. 3: typical plastic criterion for metals (standard plasticity), for granular materials (Drucker-Prager) and for amorphous silica. Arrows indicate the typical direction of flow.

The measurement of strains at the micron scale is not routine. Our approach here, in collaboration with the Laboratoire de Physico-Chimie des Matériaux Luminescents (UCB, Lyon, France) is to measure the residual silica density field by Raman microspectroscopy (Fig. 4). Some of the Raman spectral lines in silica have indeed been previously correlated to silica density.

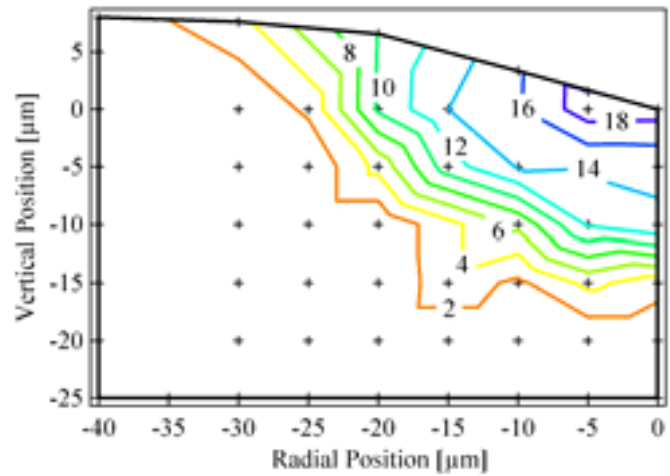

Fig. 4: residual density map obtained by Raman spectroscopy on the cross-section of an indent. The shape of the iso-density zones is a signature of the densification process specific to amorphous silica.

In this way, the residual density mappings of indented silica samples can be compared to Finite Element Calculations [Perriot]. Presently, our simulations, based on this simple plastic response, adequately accounts for the measured geometry of the plastic zone. This allows for the determination of the yield stress in a purely hydrostatic field ($P_y = 9$ GPa) and in a purely deviatoric stress field ($\tau_y = 13.5$ GPa). However these results do not completely compare with the experimental mappings: if the shape of the zone is correct, the density distribution differs. This suggests that strain hardening significantly affects the response of the materials, in analogy to the consolidation process in concrete. Adequate numerical parameters for strain hardening are presently being derived on the basis of diamond anvil cell experiments.

Of course, silica is only a test materials here. Unfortunately, for more common glasses, the Raman spectroscopic tools are inadequate. We are presently working on other markers especially using fluorescence. If the macroscopically plastic glass is still a dream, tools are being developed at least to assess the plastic contribution to the mechanical response of glass.

## *Friction control for reinforcement fibers*

In the case of reinforcement fibers, for composites, contact damage must be carefully controlled because it directly bears on the tensile strength of the fibers, one of their primary properties. But reinforcement fibers are drawn at very high velocities, typically 10 ms$^{-1}$. Such speeds are necessary to obtain the 10 μm diameter filaments wich result in the large flexural compliance and large filler/matrix interface areas and efficient reinforcement. It is well known however that shear stresses bear adversely upon contact damages. Such remarkable process velocities therefore require careful control over the friction between the fibers and the static parts.

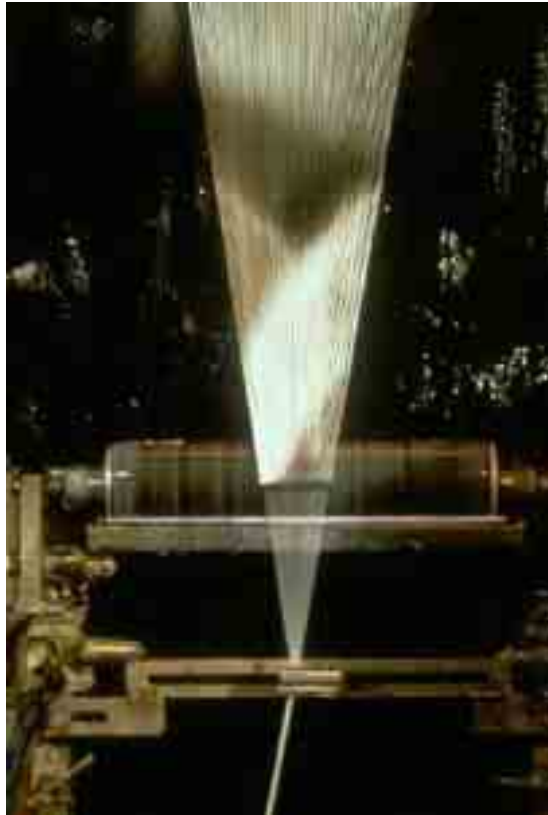

Fig. 5: a multipurpose aqueous dispersion (sizing) is applied to the glass filaments before they are pulled together to form a reinforcement fiber. Among other properties, the sizing lubricates the fiber contacts to reduce surface flaw generation. Copyright Saint-Gobain.

To control friction, an aqueous dispersion, called sizing, is applied to the reinforcement fibers as soon as they are drawn. The contact between the fibers and the first static part it meets actually aims at applying the sizing which will help protect them (Fig. 5). For some applications, the sizing contains a strongly lubricating agent, a softener, in the form of a cationic double-chain surfactant.

However, the sizing also provides several additional properties to the fibers. One of them is the ability to couple chemically to the composites polymer matrix: this is why the sizing also contains a primer, usually a silane. Silanes are small chemical species which graft to the glass surface trough their Si-OH groups and couple to organic materials through their organic groups. They are primarily used to improve the adhesion and the chemical stability of glass-organic interfaces (Fig. 6).

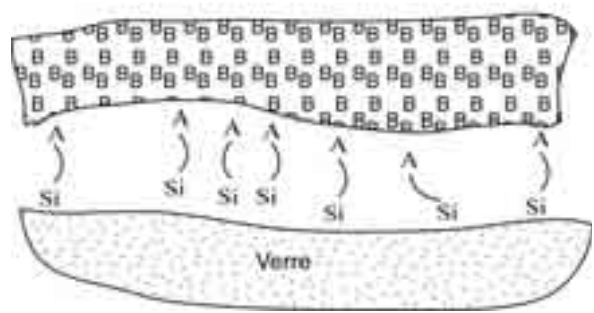

Fig. 6: schematics of the role of silane molecules as coupling agents *between glass and polymers* (cf Fig. 9)

In the sizing, therefore, one finds a mixture of these two surface active species, surfactant and silane, along with many other chemicals. Now the lubricating power of the surfactant is well-known. How does the silane interfere with it ?

By Infrared spectroscopy, we have studied the adsorption of the silane and the surfactant on their own, and the adsorption of the mixture. We observe that while the surfactant on its own adsorbs readily, due to the electrostatic interaction between the cationic head and the surface silanol groups, the adsorbed amount is severely reduced in the mixture. This in itself is not a surprise. There are several ways by which the silane may block the adsorption sites of the surfactant and reduce its affinity for the surface: one of them is the interaction between the cationic head of the surfactant and the silanol of the silane. The surprise, however, is that the lack of adsorption is not directly reflected in the friction properties.

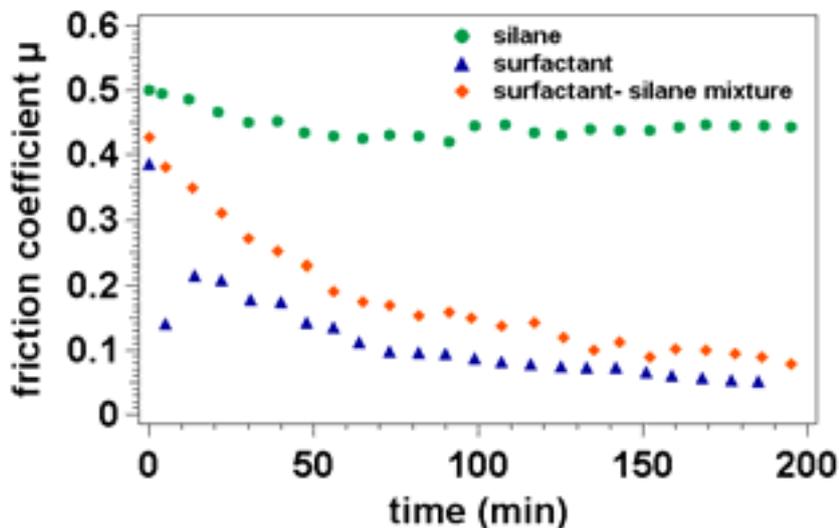

Fig. 7: low velocity (10 $\mu ms^{-1}$) friction coefficient between fused silica surfaces modified by an amino-silane, a cationic double-chain surfactant and the mixture of both. Under such contact conditions, good lubrication is obtained with the surfactant provided that sufficient organization time is allowed for. Although the silane prevents the surfactant adsorption (not shown) lubrication is still possible with the mixture, although more difficult to attain.

Indeed, we observe that the silane does not bring about lubricating properties, while the surfactant of course does (Fig. 7). The mixture, however, can lubricate the contact. The effect is simply somewhat delayed. One possible mechanism is that contact between the surfaces help induce the adsorption, because of the interaction with *two* surfaces at short separation: it is the principle of capillary condensation. Another one is that shear strains impact on the stability of the silane-surfactant complex and stabilizes the surfactant at the surface. We are still investigating this complex process [Serreau].

## *Cooperative silanes heal glass flaws*

Since surface flaws threaten the strength of glass products, a host of methods have been developed to lessen their impact. However, extremely efficient methods like ion exchange also have extremely high costs, so that for mass production, like containers (Fig. 8), glass researchers are on the look-out for cheaper - though possibly less efficient - approaches [Hand 2003].

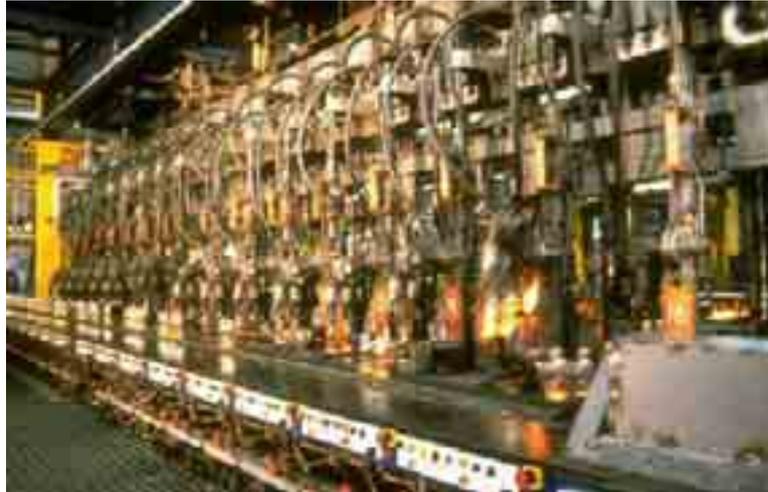
Fig. 8: Fabrication line for bottles. Copyright Saint-Gobain.

One such method has recently been developed at Saint-Gobain Recherche. Organosilanes deposited on glass surfaces on their own do not bring about reinforcement. But the key to the new process is to use *two* silane species which can react together through their organic moieties (cross-linking). Then two kinds of reactions can take place, since the silanes can also react together through their silanol groups (cross-condensation). This results in the formation of well-behaved materials.

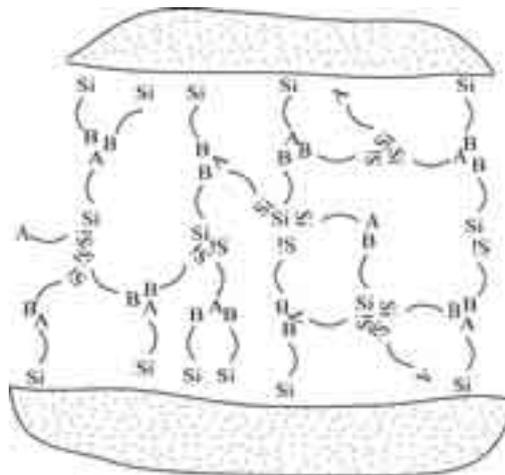
Fig. 9: silane mixture as coupling agent between *two glass surfaces* for surface flaw healing and glass strengthening (cf Fig. 6).

We have studied the strengthening power of such a mixture of epoxy- and amino-silanes [Briard 2005]. We have shown by Infra-red spectroscopy and NMR, that indeed both cross-condensation and cross-linking reactions take place. Because of the hybrid nature of the precursors, and especially the presence of reactive organic moieties, no high temperature treatment is required for the silane mixture to finally turn into a 100 nm thin solid film on the surface. A strong material is formed: by nanoindentation, we measured a 6 GPa film modulus. We also found that adhesion to glass is excellent, as expected from the original purpose of such molecules. Finally, Raman microspectroscopy has shown that the film also forms inside surface cracks. This is due to the combination of the very low viscosity of the initial aqueous dispersion and good wetting properties.

When rupture stresses are measured on calibrated indents, such a coating results in significant strengthening (75 %). On the basis of the measured mechanical properties, we proposed a strengthening mechanism based on the elastic strain of the film which bridges the crack faces. Indeed the very strong adhesion to the glass surfaces ensures that a significant fraction of the crack stresses are taken up by the elastic deformation of the film. We also point out that one of the most prominent features of such a mechanism is that the larger the cracks the better the strengthening: larger flaws are more easily healed.

The strengthening power of this system is poorer than some inorganic sol-gel or polymer films; yet it benefits from its low or even ambient temperature treatment and environmentally friendly deposition method, in addition to chemical versatility.

## *Probing the adhesion of a 10 nm thick layer: a tool for interface design*

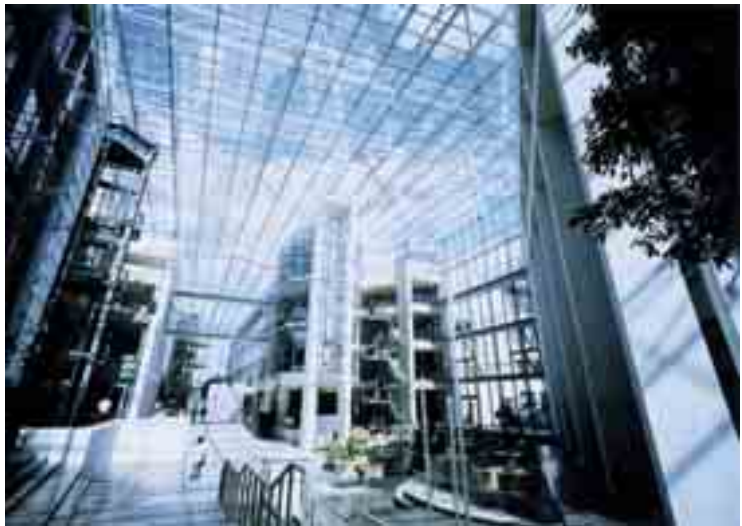

Fig. 10:
Commercial building roofed with solar control glass. Copyright Saint-Gobain.

Although glass is certainly transparent to visible light, its reflection-absorption properties are not always ideally suited over the full electromagnetic spectrum range. Correcting these properties are required for such applications as heat control (Fig. 10) where an enhanced reflection in the infra-red (IR) range is required to prevent heating by solar radiations and heat losses through IR radiations. Other properties such as anti-reflection or adaptative transmission, as presently obtained by electrochromic devices, are also highly sought for. The latter devices could result in privacy features like car windows turning opaque as one locks the doors after parking.

The standard method to – statically – tune the optical reflection-absorption is to coat the glass with stacks of extremely thin oxide and metal layers (Fig. 11). For that purpose, silver is often used for its specific dielectric response in the visible range. Silver however is a noble metal: it forms non reactive interfaces with oxides and one may expect the silver-oxide interface to be relatively weak [Didier 1994]: this will threaten the mechanical stability of the coating during processing or service.

Now partly because of the wide range of applications and party because of fundamental interest, metal/oxide interfaces have been subjected to extensive fundamental investigations in the past [Campbell 2002]. Can these results help in the design of mechanically tough optical multilayers ?

To better grasp the adhesion mechanisms in such systems, we have developed a set up to measure the adhesion of very thin multilayers on glass. It is based on the cleavage approach developed by Obreimoff in 1930. On a coated glass plate, a glass backing is glued. The sample is cleaved by inserting a blade which controls the opening of this sandwich. By measuring the opening and the crack length, the adhesion energy can be calculated. Since the opening is gradually increased by the insertion of the blade, several data points are collected on one single sample. Finally, the interface of rupture is determined by X-ray photoelectron spectroscopy of the cleaved surfaces. Although the layers are extremely thin, typically in the 10 nm range, rupture within the multilayer is perfectly interfacial and homogeneous at the atomic scale over square centimeters. Moreover, by a suitable design of test multilayers, the crack can be forced into a selected interface of rupture.

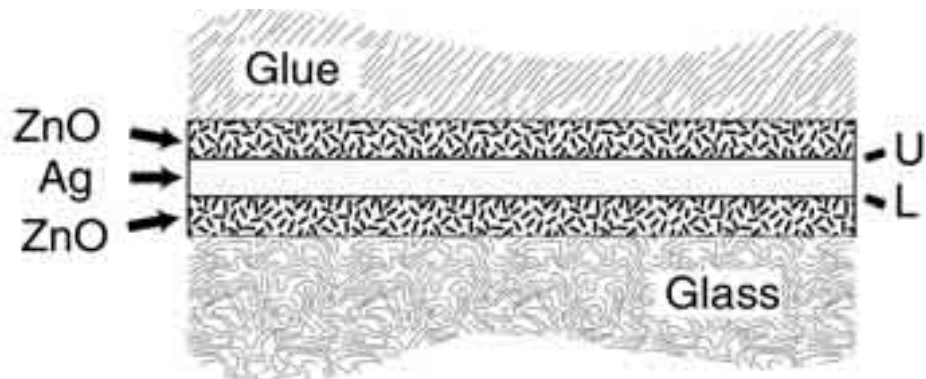

Fig. 11: Model optical multilayer with central silver layer. The L and U symbols mark the lower and upper interfaces. Reproduced from [Barthel 2005] with permission.

With silver layers, significant variations are observed for different types of oxide sublayers [Barthel 2005]. The results are generally consistent with the predictions: oxides perform better than nitrides, low gap oxides are more favourable than large gap oxides. However, some unexpected effects have also been evidenced: most noteworthy, in symmetric oxide/silver/oxide systems, we have directly evidenced that adhesion at the oxide/silver interface (denoted L on Fig. 11) differs from the silver/oxide interface (U). Such asymmetry is not expected in basic models for metal/oxide adhesion. The apparent discrepancy is due to the specific deposition conditions in PVD: because of fast film deposition rates, these two nominally identical interfaces actually have different structures. Deposition of the high surface energy element (silver) on the low surface energy element theoretically results in poor wetting: we observe in this case a weaker interface than for the converse.

Such advanced adhesion phenomena can be evidenced only by carefully conducted adhesion tests and pave the way to informed interface design for optimized multilayer mechanics. We couple this approach with structural and chemical investigation of the interfaces: we have recently demonstrated the significant role of epitaxial relations for the quality of the metal film [Sondergard 2004].

## *Porous coatings: silica materials in need of architects*

Mesoporous materials functionalities result from the presence of pores in the nanometer range. As films, these materials open new avenues for applications in different fields like electronics, catalysis, filtration and controlled release [Brinker 1990].

According to the targeted application, the necessary types of porosity differ. For low-k dielectrics, the electromagnetic excitation wavelength is large compared to the typical size of the porosity and no specific structure is needed. For filtration and controlled release, the diffusion of chemical species inside the materials must be optimized, so that the pore sizes, connectivity and tortuosity must be adjusted. As a final example, high specific area for reaction will be demanded for catalysis applications.

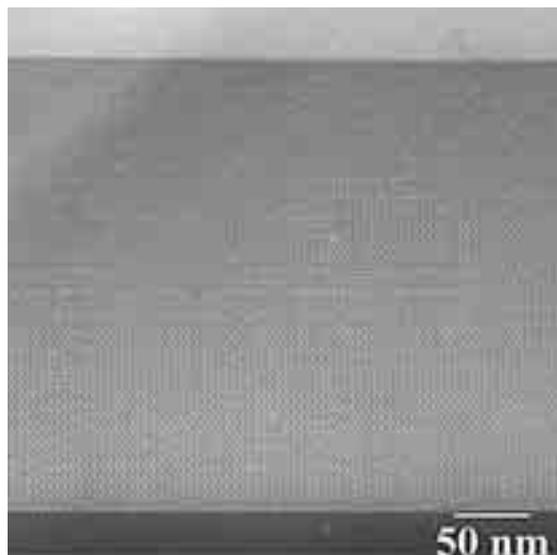
Fig. 12: TEM cross-section of a mesoporous silica film. Reprinted from S. Besson et al., J. Phys. Chem. B 104 (2000)12095 with permission.

At the same time, these functional materials require adequate mechanical properties for durability and service (Fig. 2). Now the porosity also controls the mechanical response of the materials. Can the porosity be simultaneously optimised for the system's primary function *and* the mechanical behaviour ? This is why the relation between the mechanical response and the porosity needs to be clearly understood [Roberts 2000].

For porous *films*, a full array of specific investigation techniques is required to measure their structure and understand their properties. The reason is that very often, the same material cannot be reproduced in the bulk, so that extremely sensitive experimental set-ups must be developed to tackle the problem of the very limited amount of material to be probed in thin films.

Ordered mesoporous silica films (Fig. 12) may be obtained by the combination of silane sol-gel chemistry and self-assembly of surfactants [Besson 2000]. Subsequent thermal treatments remove the templating agents by oxidation and improve the silica condensation.

We have recently studied the mechanical properties of two such films. They were synthetized with two different surfactants as templating agents. After thermal treatment, the films consist of approximately spherical pores in close packed arrangements.

The structure of the films can be measured by small angle X-ray diffraction (collaboration with P. A. Albouy (LPS, Orsay)) and the characteristics of the porosity by nitrogen adsorption isotherms directly on the films (collaboration with A. Ayral (IEM, Montpellier)) [Ayral 1998]. Transmission IR spectroscopy is used to characterize the wall material. Finally, nanoindentation probes the mechanical response of the films.

Our results indicate that films with similar pore structures and pore volume fraction, but with different pore sizes, may result in a significant 20 % difference in elastic modulus and hardness [Chemin]. This is an important result because even if the total porosity volume fraction is identical, smaller pores result in significantly larger specific area: larger pores reduce the surface available. However, the reason for the difference in mechanical properties is not entirely clear. By using different templating agents, different pore sizes are obtained, but the wall structure and especially the condensation of the silica is also be affected. The crucial idea is that structural modification at the nm scale (the mesoporosity) and at the atomic scale (the structure of the wall) often cannot be achieved independently so that a systematic approach is not feasible in practice. This is the reason why we are increasingly using IR spectroscopy to check at the atomic scale the quality of the wall materials.

These preliminary observations suggest that more research in this direction should result in a better design of porous materials through simultaneous optimization of functional porosity structure and mechanical properties. In addition to simple Young's modulus measurements, a more comprehensive mechanical characterization including toughness and ductile behaviour will be developed.

## *Conclusion*

We have shown how material science and mechanics couple for better understanding of the response of glass and functionalised glass surfaces.

The intrinsic behaviour of glass, because of its amorphous nature, is highly unusual, and exhibits non standard ductile behaviour. As a prototype, we have studied amorphous silica because it is particularly anomalous, with large density variations, and also amenable to local mechanical characterization through Raman microspectroscopy. Comparing experimental density maps with Finite Element calculations with reasonable constitutive laws, we provide more insight into silica ductile behaviour. Understanding ductility in silicate glasses may be lead to a better control over their brittleness.

Another challenge is provided by film-functionalised glass surfaces: the final mechanical behaviour of the surface and the stability of the film will result from the interplay of a rich variety of phenomena (friction, adhesion, intrinsic mechanical properties). These material properties must be characterized and understood at the scale of the surface or the scale of the film. For that purpose, due to minute film thicknesses, specific experimental devices must be developed and adequate models devised.

## *Acknowledgements*

The authors are grateful to many researchers and technicians at Saint-Gobain Recherche and at Vetrotex International for their help, discussions and suggestions, most notably among the mechanics, thin films and composites departments.## **References:**

A. Ayral et al., J. Mater. Sci. Lett. 17 (1998) 883.
E. Barthel et al., Thin Solid Films 473/2 (2005) 272.
S. Besson et al. J. Mater. Chem. 10 (2000) 1331.
R. Briard et al., J. Non-Cryst. Sol. 351 (2005) 323.
C. Brinker and G. Scherer, "Sol-Gel Science" Academic Press, San Diego, 1990.
C. Campbell and D. Starr, J. Am. Chem. Soc. 124 (2002) 9212.
N. Chemin et al., in preparation.
F. Didier and J. Jupille, Surface Science 314 (1994) 378.
B. Derjaguin, Kolloid-Z. **69** (1934) 155.
F. Ernsberger, J. Am. Ceram. Soc. 51 (1968) 545.
A. A. Griffith, Phil. Trans. Roy. Soc. A 221 (1921) 163.
R. Hand et al., J. Non-Cryst. Sol. 315 (2003) 276.
J. Lambropoulos, J. Am. Ceram. Soc. 79 (1996) 1441.
J. W. Obreimoff, Proc. Roy. Soc. 127 (1930) 290.
A. Perriot et al., to appear in J. Am. Ceram. Soc.
Y. W. Rhee et al., J. Am. Ceram. Soc. 84 (2001) 561.
A. Roberts and E. Garboczi, J. Am. Ceram. Soc. 83 (2000) 3041.
L. Serreau et al., in preparation.
E. Sondergard et al., Surface Science 559 (2004) 131.

## *SVI*

Surface du Verre et Interfaces, a CNRS/Saint-Gobain research unit.

The mission of "Surface du Verre et Interfaces" (SVI) is to perform fundamental research relevant to Saint-Gobain processes and products. It thus contributes to the scientific exchanges between public and private research so as to simultaneously promote industrial innovation and basic knowledge. For that purpose, and to optimize contacts, the laboratory is located in the Saint-Gobain Recherche premises, one of the three main Saint-Gobain research centers. At the same time, the unit develops close connections with public research laboratories in France and also abroad.

In the near future, the lab, while still developing its expertise in surface mechanics, will focus on structured coatings, with emphasis on PVD, non-silicate sol-gels and polymers.

http://www.saint-gobain-recherche.com/svi/